\documentclass{article}
\usepackage[tbtags]{amsmath}
\usepackage{amsmath}
\usepackage[portrait, margin=1in]{geometry}
\usepackage{amssymb}
\usepackage{booktabs}
\usepackage{tikz}
\usetikzlibrary{positioning}
\usepackage{bbm}
\usepackage{amsthm}
\usepackage{authblk}
\usepackage{float}
\usepackage[toc,page]{appendix}
\usepackage{bm}
\usepackage[bottom]{footmisc}
\usepackage{tikz}
\usetikzlibrary{bayesnet}
\usetikzlibrary{arrows}
\usepackage{color}
\usepackage{graphicx}
\usepackage{caption}
\usepackage{subcaption}
\usetikzlibrary{backgrounds}
\usepackage{upgreek}
\usepackage{graphicx}
\usepackage{mathtools}
\usepackage[T1]{fontenc}
\usepackage{lipsum}   % for filler text
\usepackage{setspace} % for \onehalfspacing and \singlespacing macros
\usepackage{algorithm}
\usepackage{algpseudocode}
\usetikzlibrary{positioning}
\usepackage[utf8]{inputenc}
\usepackage{algorithm,algcompatible,amssymb,amsmath}
\usepackage{algorithm}
\usepackage{algpseudocode}

\algnewcommand\algorithmicto{\textbf{to}}
\algnewcommand\RETURN{\State \textbf{return} }

\title{Bayesian Estimation of Corporate Default Spreads}
\author{Maksim Papenkov$^{1,2,}$\footnote{Corresponding Author : mp3827@columbia.edu} $\;\&$ Beau Robinette$^{1}$}
\date{%
    $^1$O'Shaughnessy Asset Management\\%
    $^2$Columbia University, Department of Computer Science\\[5ex]%
    \today
}

\begin{document}
\maketitle
\vspace{-0.6cm}
\begin{abstract}
    Risk-averse investors often wish to exclude stocks from their portfolios that bear high \textbf{credit risk}, which is a measure of a firm's likelihood of bankruptcy. This risk is commonly estimated by constructing signals from quarterly accounting items, such as debt and income volatility. While such information may provide a rich description of a firm's credit risk, the low-frequency with which the data is released implies that investors may be operating with outdated information. In this paper we circumvent this problem by developing a high-frequency credit risk proxy via corporate default spreads which are estimated from daily bond price data. We accomplish this by adapting classic yield curve estimation methods to a corporate bond setting, leveraging advances in Bayesian estimation to ensure higher model stability when working with small sample data which also allows us to directly model the uncertainty of our predictions. 
\end{abstract}

\vspace{0.2cm}

\newpage
\section{Research Motivation}

\subsection{Problem} 
A single extreme credit event can significantly erode an investor's accumulated wealth. While stocks for credit risky firms offer a high potential upside, for investors with a low tolerance for risk it may be desirable to exclude such firms from their portfolios entirely. Without sufficient data and an appropriate credit risk model, identifying risky firms is generally difficult - though an emerging literature has offered several viable model frameworks \cite{merton1974pricing, altman1968financial, leland1996optimal}. Utilizing a robust credit model, a risk-averse investor has the ability to avoid stocks with high credit risk, potentially providing a significant reduction to the drawdown risk of their portfolio. 

\subsection{Existing Solution (and its Limitations)} 
The standard approach towards modeling issuer credit risk is with a \textbf{structural credit risk model} that predicts some exogenous event (such as a bankruptcy) with a factor model \cite{leland_structural_model}. Often, the factors used within such a model are derived from quarterly financial statements, primarily focusing on a firm's quantity of debt and its estimated ability to repay it. While such models are generally very effective in practice, they have two critical limitations : 
\vspace{-0.3cm}
\begin{itemize}
    \item \textbf{High-Latency} : Financial statements are prepared by accountants and take several weeks (or months) to process, and thus do not always represent the most recent information. 
    \item \textbf{Low-Frequency} : Per SEC rules, financial statements are only issued quarterly, and if some high-impact even occurs in between financial statements, the model would become inconsistent with reality. 
\end{itemize}
\vspace{-0.3cm}
We can circumvent these issues by pivoting our attention away from quarterly financials to daily bond price data, which is directly influenced by the market's perception around a firm's ability to repay its debts. 

\subsection{New Solution (and its Value Proposition)}

We introduce the \textbf{Bayesian Default Spread} which adapts classic methods for yield curve estimation to a corporate bond setting. Corporate bond prices are generally much more difficult to work with than government bonds - as the sample sizes are much smaller, the assets tend to be less liquid, and the data tends to be more noisy. We address these issues with Bayesian estimation, which allows us to regularize our parameters in a way that ensures stability despite the data limitations. 

We structure this paper as a full methodological tutorial, beginning with first principals and gradually building up to the full Bayesian method. This new method offers the following benefits : 
\vspace{-0.3cm}
\begin{itemize}
    \item \textbf{Low-Latency} : since corporate bonds are traded in large markets, their prices quickly adjust to reflect newly available information on a firm. 
    \item \textbf{High-Frequency} : for many firms, their bonds are traded many times within a quarter, sometimes daily, providing a full time-series of changes throughout a year. 
\end{itemize}
\vspace{-0.3cm}
A clear limitation of our method is that not all firms issue bonds, limiting the utility of the Bayesian Default Spread as a credit risk signal to only large-cap firms. However, fixed income markets have significantly expanded in the last decade, and as the number of traded bonds has increased the viability of this method has improved in recent years. 

\newpage
\section{Principals for Bond Pricing}
Before delving into regression architectures, we'll first discuss the necessary prerequisite math. Formally, a \textbf{bond} is a contract providing a deterministic sequence of $M$-many cash payments, such that each payment is denoted as $c_t$, represented in currency. These payments are temporally-indexed by $\mathcal T \subset \mathbb R_+$ : 
\begin{equation}\label{bond definition}
    \mathcal B = \{c_t\}_\mathcal T = \{c_{t_1}, c_{t_2}, \dots, c_{t_M}\}
\end{equation}
A bond's \textbf{principal} is the value of the terminal payment and its \textbf{coupon} is the value of the interim payment. The coupon is quoted in terms of a \textbf{coupon rate}, which is a percentage of the principal, and a final coupon is paid alongside the principal. The total contract length is the bond's \textbf{maturity}, which starts on the bond's \textbf{issue date} and ends on the bond's \textbf{maturity date}. The \textbf{coupon frequency} is the number of coupon payments made per year. Beyond this simple structure, there also exist many flavors of \emph{exotic} bonds with additional complexity, such as variable rates and optionality, though we exclude them from our scope. 

For our model we only consider \textbf{vanilla} bonds - senior debt with fixed coupon rates,  no optionality, and no convertibility. Simply put, conditional on no default, a vanilla bond has a fully deterministic cash flow. For example, consider some particular two-year maturity vanilla bond $\mathcal B^*$ with a principal of $\$1000$, a coupon rate of $5\%$ paid semi-annually ($\$25$ every six months), which has the following cash flow sequence : 
\begin{equation}
    \mathcal B^* = \{c_{0.5},\; c_{1},\; c_{1.5},\; c_{2}\} = \{\$25,\; \$25,\; \$25,\; \$1025\}
\end{equation}
In general, the price of a bond is not simply equal to the sum of cash payments $\sum c_t$, but rather it is influenced by a set of external factors. For this, we require asset pricing theory. 
\vspace{-0.3cm}
\subsection{Necessary Assumptions}\label{axioms}
\vspace{-0.1cm}
To obtain a pricing model for a bond, we require a theoretical framework with necessary assumptions. First, we define a \textbf{discount function} as a mapping from time to the unit interval such that : 
\begin{equation}
    \text d(t) \coloneqq  
    \begin{cases} 
        \text d(t) = 1 & \text{if } t = 0 \\
        \text d(t) \in [0, 1) & \text{if } t \in \mathbb R_+
    \end{cases}
\end{equation} 
In practice, this function is \emph{latent} and must be inferred from data, thus we can consider $\text d(t)$ as a stochastic process, and assume that \emph{uncertainty increases monotonically over time} : 
\begin{equation}
        \delta > 0 : \mathbb V\big[\text d(t)\big] < \mathbb V\big[\text d(t + \delta)\big]
    \end{equation}
For such a function to exist within a market, we must assume \emph{a strictly positive risk-free rate exists} :
\begin{equation}
        \exists\; r_\text{risk-free} : r_\text{risk-free} > 0
\end{equation}
Thus, for every bond we assume the existance of two separate discount functions : 
\vspace{-0.3cm}
\begin{enumerate}
    \item \textbf{Time Value Discounting} : \emph{a dollar today is worth strictly less than a dollar tomorrow}. Since we assume a strictly positive risk-free rate exists, then an investor is able to invest a dollar today and obtain strictly greater than a dollar back in the future. Such a discount function is constant across all issuers in a market, and can never be eliminated. We'll denote this as $\text d_\text{time}(t)$. 
    \item \textbf{Default Risk Discounting} : \emph{a risk of failing to receive payment decreases the price of a bond}. An investor is only willing to accept additional risk for additional compensation, and higher risk requires greater discounting. Such a discount function is unique to each issuer. We'll denote this as $\text d_\text{risk}$(t). 
\end{enumerate}
\vspace{-0.3cm}
We assume the risk-free rate corresponds to the US Treasury, and thus assume it has no risk of default. We define the following \emph{degenerate} default risk discount function for the US Treasury as : 
\begin{equation}
    \forall t\in \mathbb R_+ : \text d^*_\text{risk}(t) = 1
\end{equation}
\textbf{Big Idea} : The core purpose of this paper is to describe a general method for inferring $\text d_\text{risk}(t)$ for any corporate issuer from observed bond price data, which will act as an ordinal bankruptcy prediction proxy signal that can be used for stock selection and portfolio construction. 

\subsection{Bond Pricing via Present Value}

We can now bring together the cash flow sequence associated with a bond and a discount function to obtain a fair price for a bond. Utilizing the assumptions from the previous section, we define the \textbf{composite discount function} for a particular issuer as : 
\begin{align}
    \text d_\text{total}(t) = \text d_\text{time}(t) \cdot \text d_\text{risk}(t)
\end{align}
The price $p$ of bond $\mathcal B$ is defined via the \textbf{present value} function, which is simply the sum of all future discounted cash flows. Notice that this is upper-bounded by the sum of the cash flows  : 
\begin{equation}
    p = \text{PV}\big(\mathcal B, \text d_\text{total} \big) =  \sum_{t\in \tau} c_{t} \cdot \text d_\text{total}(t) \leq \sum_{t\in\tau} c_t
\end{equation}
\vspace{-0.7cm}

For convenient computation, we represent this linear algebraically by constructing the following vectors : 
\vspace{-0.3cm}
\begin{itemize}
    \item \textbf{Cash Flow Vector} : $\mathbf c = \big[c_{t_1}, c_{t_2}, \dots, c_{t_M}\big]\in \mathbb R_+^{M}$
    \item \textbf{Discount Factor Vector} : $\mathbf d = \big[\text d(t_1), \text d(t_2), \dots, \text d(t_M)\big] \in [0,1]^{M}$
\end{itemize}
\vspace{-0.3cm}
Thus, we the present value function reduces to the following inner product : 
\begin{equation}\label{present_value}
    p = \text{PV}(\mathcal B, \text d_\text{total}) = \mathbf c^\text T \mathbf d \in \mathbb R_+
\end{equation}
We emphasize that the price of any bond incorporates \emph{both} time value and default risk discounting simultaneously, and we must work to disentangle the two. Fortunately, since we assume that US Treasury bonds are risk-free, this disentanglement will be surprisingly easy to accomplish.

\subsection{Yield Curves and Default Spread}

While the discount function is mathematically convenient, it is more natural to compare firms in terms of \textbf{yield curves}, which are theoretical interest rates on zero-coupon bonds maturing at time $t$, defined by the following isomorphic transform : 
\begin{equation}
    \text y(t) = -\log\big[\text d(t)\big] \cdot t^{-1} \iff \text d(t) = \exp\big[- \text y(t) \cdot t\big]
\end{equation}
Since we assume a positive risk-free rate exists, this implies that we can always construct a well-defined yield curve on US Treasuries. A \textbf{default spread}, sometimes called a \emph{credit spread} is the difference between an issuer's total yield and the US Treasury yield :
\begin{align}
    \text s\big(t\big) 
    &= \text y_\text{total}\big(t\big) - \text y_\text{Treasury}\big(t\big) \\[5pt]
    &= -\log\big[\text d_\text{total}\big(t\big)] \cdot t^{-1} + \log\big[\text d_\text{time}\big(t\big)\big] \cdot t^{-1} \\[5pt]
    &= -\log\big[\text d_\text{risk}\big(t\big)] \cdot t^{-1} -\log\big[\text d_\text{time}\big(t\big)] \cdot t^{-1} + \log\big[\text d_\text{time}\big(t\big)\big] \cdot t^{-1}\\[5pt]
    &= -\log\big[\text d_\text{risk}\big(t\big)] \cdot t^{-1} 
\end{align}
The above derivation clearly demonstrates that the default spread is simply a monotonic function of an issuer's \emph{risk discount function}, providing an ordinal signal for ranking firms on the basis of credit-riskiness. We can estimate total default risk as the following integral over the default spread : 
\begin{equation}
    \text{risk}(T) = \int_0^T \text s(t)\; dt
\end{equation}
\textbf{Summary in simpler terms : } in an approximately efficient market, a firm with high bankruptcy risk has bond prices that reflects this risk. Although we cannot directly observe a firm's discount function, we can however estimate a default spread from data. In practice, this is good enough. 

\newpage
\section{Data}
While classical credit risk methods use quarterly fundamentals, we rely \emph{only} on bond price data. It is important for the reader to keep in mind that bond markets have grown significantly in the past several years, and coverage for corporate bonds is gradually increasing. This makes proper backtesting particularly difficult, as the nature of bond markets now is much different from what it was even a decade ago. For our results section we will thus only focus on a few sub-samples for illustrative purposes. 

\subsection{Sourcing}

We rely on the following data sources to construct a signal : 
\vspace{-0.3cm}
\begin{itemize}
    \item \textbf{TRACE Bond Prices} \cite{trace_finra} : this includes daily closing prices for all publicly traded bonds, though for our purpose we are only interesed in corporate securities. 
    \item \textbf{FISD Issue Data} \cite{fisd_mergent} : this includes descriptive data, such as issue and maturity dates, coupon rates and frequencies, and flags for things like optionality and convertibility. 
    \item \textbf{Treasury Yield Curve} : we use this to convert a corporate yield into a corporate default spread. 
\end{itemize}
\vspace{-0.3cm}
As previously mentioned, we constrain our analysis to a subset of \emph{vanilla} bonds, which are the simplest class of bonds with no random structure to their payout. 

\subsection{On-the-Run Bonds}

In general, issuers often have multiple bonds in a market for a single maturity, which tend to have overlapping payouts. An empirical phenomenon that we must accommodate is the fact that when a firm issues a bond for a maturity that another issued bond already has, the liquidity of the older bond significantly falls. Thus, we only consider \textbf{on-the-run} bonds, which are the most recent issue of a particular maturity for an issuer. 

For example, suppose a firm issued a 30-year bond last year, and another 30-year bond today. Only the bond issued \emph{today} would be considered on-the-run, and the other would be excluded from our dataset. This is also something that must be accounted for when working with government bonds. 

\subsection{Dirty Prices}

This is really just a note on data pre-processing. In practice, investors require compensation for selling a bond in-between coupon payments. We define \textbf{accrued interest} as the value of a partial coupon : 
Let $t_\text{next}$ be the time until the next coupon payment, let $t_\text{coupon}$ be the coupon period, and let $c_\text{next}$ be the cash value of the next coupon. Thus, \textbf{accrued interest} is simply defined as 
\begin{align}
    a = c_\text{next} \cdot \frac{t_\text{next}}{t_\text{coupon}}
\end{align}
The \textbf{dirty price} is simply the sum of the \emph{clean price} and the \emph{accrued interest} : 
\begin{align}
    p_\text{dirty} = p + a
\end{align}
To avoid a notational headache, for the remainder of this paper, $p$ will refer only to the \emph{dirty price} of a bond. 

\section{Regressions for Default Spread Modeling}

\subsection{Vasicek Exponential Spline}\label{vasicek_spline}
We estimate an issuer's total discount function with the following \textbf{linear basis function model}\footnote{In case this analogy is helpful to the reader, we can think of this framework as similar to a polynomial regression, in which we estimate $f(x) = \boldsymbol\upphi_x^\text T \boldsymbol\beta \in \mathbb R$ with $\boldsymbol\upphi_x = [x, x^2, x^3, \dots, x^K]$. In both cases, we approximate a complicated function as a linear combination of simple functions, though a polynomial regression maps to $\mathbb R$ while our discount model only maps to $[0,1]$.}, in which we use a linear combination of simple basis functions to approximate our true latent function. We specifically select $\upphi_{\alpha,k}:\mathbb R_+ \rightarrow [0,1]$ to be a well defined discount function  : 
\vspace{-0.2cm}
\begin{equation}\label{vasicek_linear_basis_model}
    \text d^*_{\beta}(t) = \sum_{k=1}^K \upphi_{\alpha, k}(t) \cdot \beta_k = \boldsymbol \upphi_t^\text T \boldsymbol\beta  \in \big[0,1\big]
\end{equation}
\vspace{-0.7cm}

In this model, we represent a single discount factor for a particular payout time $t$ as the inner product between latent $\boldsymbol\beta \in \mathbb R^K$ (which we will estimate via a linear regression) and \textbf{basis function vector} $\boldsymbol\upphi_t$ as :  
\begin{align}\label{phi_vector}
    \boldsymbol\upphi_t = \big[\upphi_{\alpha, 1}(t), \upphi_{\alpha, 2}(t), \dots, \upphi_{\alpha,K}(t)\big]\in \big[0,1\big]^K
\end{align}
There exist several reasonable choices for $\upphi$ in the literature, such as polynomials, fourier series and exponential splines \cite{bank_of_canada_paper} \cite{mles_paper}. We select the \textbf{Vasicek basis} \cite{vasicek_spline} due to its flexibility.  : 
\begin{equation}
    \upphi_{\alpha, k}(t) = \exp(-\alpha \cdot k \cdot t\big) \in \big[0,1\big]
\end{equation}
Notice that with this basis $\phi_{\alpha,k}(0) = 1$. To avoid a pathological basis function model in which $d(0) \neq 1$, we constrain $\|\boldsymbol\beta\|_1 = 1$. We'll handle the details of this constraint in section \ref{numerical_consid_beta}.
To price a bond using this discount function, we construct a Vasicek \emph{discount factor vector} for a bond with $M$-many cash flows indexed by $\mathcal T = \{t_1, t_2, \dots, t_M\}$ as : 
\begin{equation}
    \mathbf d^* = \big[\text d_\beta^*(t_1), \text d_\beta^*(t_2), \dots, \text d_\beta^*(t_M)\big] = \boldsymbol\Phi^\text T \boldsymbol\beta \in \big[0,1\big]^M
\end{equation}
 Here we define \textbf{basis factor matrix} $\boldsymbol\Phi$, which is an $M$-dimensional extension of Equation \ref{phi_vector}. This is simply a matrix of basis discount factors aligned with the payout times of cash flows for our bond.  To simplify our task, we fix $\alpha=0.05$ and suppress it in our notation  : 
\begin{equation}
    \boldsymbol\Phi = 
    \begin{bmatrix}
    \upphi_{1}\big(t_1\big) &  \cdots & \upphi_{K}\big(t_1\big)  \\
    \vdots   & \upphi_{k}\big(t_m\big)  & \vdots  \\
    \upphi_{1}\big(t_{M_n}\big) &  \cdots & \upphi_{K}\big(t_{M_n}\big) 
    \end{bmatrix}
    \in \big[0,1\big]^{{M} \times K}
\end{equation}
Thus, the price of a bond discounted by the Vasicek discount function is simply : 
\begin{equation}
    p = \mathbf c^\text T \mathbf d^* = \mathbf c^\text T \boldsymbol\Phi^\text T \boldsymbol\beta \in \mathbb R_+
\end{equation}
We emphasize that $p, \mathbf c$ and $\boldsymbol\Phi$ are known ahead of time, while only $\boldsymbol\beta$ is latent - thus, a linear regression naturally follows. For our regression, we construct the following \textbf{basis price vector} as : 
\begin{equation}
    \mathbf b = \mathbf c^\text T \boldsymbol\Phi = \big[\text{PV}( \mathcal B,\upphi_{1}\big), \dots, \text{PV}( \mathcal B,\upphi_{K})\big]\in \mathbb R^K
\end{equation}
This naturally extends to the following \textbf{basis price matrix} for $N$-many bonds : 
\begin{equation}
    \mathbf B = 
     \begin{bmatrix}
     \text{PV}( \mathcal B_1,\upphi_{1}\big)&  \dots&  \text{PV}( \mathcal B_1,\upphi_{K})
     \\
     \vdots   & \text{PV}( \mathcal B_n,\upphi_{k}\big)  & \vdots  \\
     \text{PV}( \mathcal B_N,\upphi_{1}\big)&  \dots&  \text{PV}( \mathcal B_N,\upphi_{K})
     \end{bmatrix}
     \in \mathbb R^{N \times K}
\end{equation}
Thus, given a set of $N$-many bonds, we define a linear regression of the form: 
\begin{equation}
    p_n = \mathbf b_n^\text T \boldsymbol\beta + \varepsilon_n 
\end{equation}
To make notation more consistent with what the reader may be used to, for the remainder of the paper we will use the convention $y_n \coloneqq p_n$ and $\mathbf x_n\coloneqq \mathbf b_n$. Now that we have defined the data structure that we will use to fit our regression, we must consider several numerical subtleties. Similar architectures have been discussed in the context of goverment bonds in \cite{australian_yield_curve,bank_of_canada_paper,bank_of_italy_paper, mles_paper, nelson_siegel_1987}.
\subsection{Numerical Considerations for Estimating $\boldsymbol\beta$} \label{numerical_consid_beta}

As a naive initial case, we can consider an estimate of $\boldsymbol\beta$ via \textbf{ordinary least squares (OLS)}, which is a convenient closed-form convex optimization approach : 
\begin{align}
   \hat{\boldsymbol \beta}_\text{OLS}
   & =\arg\min_{\boldsymbol\beta}\Big\{ \big\|\mathbf y - \mathbf X^\text T \boldsymbol\beta\big\|_2^2 \Big\}
   \\
   &=\big(\mathbf X^\text T \mathbf X\big)^{-1}\big(\mathbf X^\text T\mathbf y\big) \in \mathbb R^K
\end{align}
While OLS is a reasonable starting point for fitting a Vasicek exponential spline, unfortunately it isn't sufficiently structured to model corporate bonds. We must generalize our approach and account for three additional considerations that we care about in practical implemenation.  

\textbf{Issue 1} : \emph{if $\sum \beta_k \neq 1$, then $\mathrm{d}(0) \neq 1$, producing a pathological discount function.} Simply, it does not make sense to discount $\$1$ today to anything other than $\$1$. The natural solution to this issue is to fit a constrained regression with $\|\boldsymbol\beta\|=1$, but equality constraints are a hassle to work with. Instead, we map our constrained problem in $\mathbb R^K$ to an unconstrained problem in $\mathbb R^{K-1}$ with the following data transform, allowing us to stay within a least-squares framework, defining $\mathbf X_k$ as the $k$-th column of $\mathbf X$ (see proof in Appendix \ref{constrained_reg_proof}) : 
\begin{align}
    \tilde{\mathbf y} &= \big[\mathbf y - \mathbf X_K\big] \in \mathbb R^N
    \\[5pt]
    \tilde{\mathbf X} &= \big[\mathbf X_1 - \mathbf X_K, \dots, \mathbf X_{K-1} - \mathbf X_K\big] \in \mathbb R^{N \times (K-1)}\\[5pt]
    \tilde{\boldsymbol\beta} &= \big[\beta_1, \dots, \beta_{K-1}] \in \mathbb R^{K-1}
\end{align}
\textbf{Issue 2} : \emph{some observations are more reliable than others}. Per Axiom 3 in Section \ref{axioms}, it follows that the price uncertainty of a bond is proportionate to its total term. Thus, when fitting a Vasicek spline, it is not appropriate to weight all observations equally in the objective function. For each bond, we define a weighting factor inversely proportion to its term as $w_i \propto M_n$, with which we construct diagonal \emph{weight matrix} $\mathbf W\in\mathbb R^{N\times N}$, which penalizes less-reliable observations via \textbf{weighted least squares (WLS)} : 
\begin{align}
   \hat{\tilde{\boldsymbol \beta}}_\text{WLS}
   & =\arg\min_{\boldsymbol\beta}\Big\{ \big\|\sqrt{\mathbf W} \big(\tilde{\mathbf y} -\tilde{ \mathbf X}^\text T \tilde{\boldsymbol\beta})\big\|_2^2 \Big\}
   \\
   &=\big(\tilde{\mathbf X}^\text T \mathbf W\tilde{\mathbf X}\big)^{-1}\big(\tilde{\mathbf X}^\text T\mathbf W\tilde{\mathbf y}\big)\in\mathbb R^{K-1}
\end{align}
\textbf{Issue 3} : \emph{for some corporate issuers, $N < (K-1)$}. While the above WLS method is perfectly suitable for estimating the Treasury discount function, since for Treasuries we often have $N>10$, things break down for issuers with $N<(K-1)$ as $\tilde{\mathbf X}^\text T \mathbf W \tilde{\mathbf X}$ becomes uninvertible. Rather than restrict the expressivity of our model by decreasing $K$, we can instead resolve this issue by applying L2-regularization (``ridge regression'') on $\boldsymbol\beta$, which produces an always-invertible solution via \textbf{ridge weighted least squares (RWLS)} : 
\begin{align}\label{ridge_regression}
   \hat{\tilde{\boldsymbol \beta}}_\text{RWLS}
   & =\arg\min_{\boldsymbol\beta}\Big\{ \big\|\sqrt{\mathbf W} \big(\tilde{\mathbf y} - \tilde{\mathbf X}^\text T \tilde{\boldsymbol\beta})\big\|_2^2 + \lambda \big\|\tilde{\boldsymbol\beta}\big\|_2^2\Big\}
   \\
   &=\big(\tilde{\mathbf X}^\text T \mathbf W\tilde{\mathbf X} + \lambda \mathbf I\big)^{-1}\big(\tilde{\mathbf X}^\text T\mathbf W\tilde{\mathbf y}\big) \in \mathbb R^{D-1}
\end{align}
\textbf{Summary} : To simplify this architecture and avoid the clunky \emph{tildes} above, we introduce the following transform, utilizing $\mathbf p$ and $\mathbf B$ from Section \ref{vasicek_spline}, and retain this notation for the remainder of the paper. We use $\mathbf B_k$ to denote the $k$-th \emph{column} of matrix $\mathbf B$ : 
\begin{align}
    \mathbf W &\coloneqq \text{diag}\big([w_1, \dots, w_N]\big) \in \mathbb R^{N \times N} \text{ s.t. } w_n = f(M_n)
    \\[5pt]
    \label{new_y}\mathbf y &\coloneqq \sqrt{\mathbf W} \big( \mathbf p - \mathbf B_K\big) \in \mathbb R_+^N
    \\[5pt]
    \label{new_X}\mathbf X &\coloneqq \sqrt{\mathbf W} \big[\mathbf B_1 - \mathbf B_K, \dots, \mathbf B_{K-1} - \mathbf B_K\big] \in \mathbb R_+^{N\times(K-1)}
\end{align}
We have not yet spoken on a reasonable method for selecting $\lambda$, though in practice an optimal value may be discovered through cross-validation. Alternatively, in the next section we introduce a broader approach to regularization via \emph{Bayesian inference} that provides us with a general framework to incorporate additional constraints. Eventually, we will further expand this to a temporal model using state-space methods. 

\newpage
\subsection{Estimating $\boldsymbol\beta$ via Bayesian Inference}

We generalize the regularization on our regression by introducing \textbf{Bayes Theorem} : 
\begin{align}
    \underbrace{\mathbb P\big(\boldsymbol\theta\mid \mathbf X \big)}_\text{posterior} \propto \underbrace{\mathbb P\big(\mathbf X \mid\boldsymbol\theta\big)}_\text{likelihood} \;\; \underbrace{\mathbb P\big(\boldsymbol\theta\big)}_\text{prior}
\end{align}
The \textbf{likelihood} defines the relationship between our data and the model parameters (similar to the OLS objective), the \textbf{prior} defines all constraints on our parameters (similar to the L2-regularizer), and the \textbf{posterior} ties it all together. The primary philosophical shift here is that we will now think of our parameters as \emph{random variables} with corresponding probability distributions, which allows us to easily construct predictive confidence intervals that incorporate all of the uncertainty embedded in a model. 

We can extend the \emph{ridge regression} (Equation \ref{ridge_regression}) by introducing the \textbf{conjugate Bayesian regression} (utilizing $\mathbf y$ and $\mathbf X$ defined in Equations \ref{new_y} and \ref{new_X}), which has the following generative model : \begin{align}
    y_n &\sim \mathcal N\big(y_n \mid \mathbf x_n^\text T \boldsymbol\beta, \sigma^2\big) \\
    \boldsymbol\beta\mid \sigma^2 &\sim \mathcal N\big(\boldsymbol\beta \mid \boldsymbol\mu, \sigma^2  \boldsymbol\Lambda\big)\\
    \sigma^2 &\sim \mathcal G^{-1}\big(\sigma^2 \mid \alpha, \gamma\big)
\end{align}
This model requires $\boldsymbol\mu\in\mathbb R^{(K-1)}$, positive semi-definite $\boldsymbol\Lambda\in\mathbb R^{(K-1),(K-1)}$, and $\alpha,\gamma \in \mathbb R_+$. To simplify, we use a \textbf{Normal Inverse-Gamma} distribution on $\boldsymbol\theta = \{\boldsymbol\beta,\sigma^2\}$ as : 
\begin{align}
    \mathbb P\big(\boldsymbol\theta\big) = \mathbb P\big(\boldsymbol\beta, \sigma^2\big) =\mathbb P\big(\boldsymbol\beta\mid \sigma^2\big) \cdot \mathbb P\big(\sigma^2\big) = \mathcal {NG}^{-1}\big(\boldsymbol\beta, \sigma^2\mid\boldsymbol\mu,\boldsymbol\Lambda,\alpha,\gamma\big)
\end{align}
Using the notation that $\theta_0$ is a prior parameter and $\theta_*$ is a posterior parameter, we have posterior : 
\begin{align}
    \mathbb P\big(\boldsymbol\theta \mid\mathbf X\big) &=\mathcal {NG}^{-1}\big(\boldsymbol\beta, \sigma^2\mid\boldsymbol\mu_*, \boldsymbol\Lambda_*, a_*, b_*\big)
    \\[5pt]
    \boldsymbol\Lambda_* \label{eq_45}
    &= \big(\mathbf X^\text T \mathbf X + \boldsymbol\Lambda_0^{-1}\big)^{-1}
    \\[5pt]
    \boldsymbol\mu_* 
    &= \boldsymbol\Lambda_* \big(\mathbf X^\text T\mathbf y + \boldsymbol\Lambda_0\boldsymbol^{-1}\mu_0\big)
    \\
    \alpha_* \label{eq_47}
    &= \alpha_0 + \frac{N}{2}
    \\
    \gamma_* \label{eq_48}
    &= \gamma_0 + \frac{1}{2}\big[\mathbf y^\text T\mathbf y + \boldsymbol\mu^\text T_0 \boldsymbol\Lambda_0^{-1}\boldsymbol\mu_0
    -\boldsymbol\mu_*^\text T \boldsymbol\Lambda_*^{-1}\boldsymbol\mu_*\big]
\end{align}
Equipped with these posterior parameters, we construct the predictive distribution for $\mathbf x_\text{new} \in \mathbb R^{K-1}$ as : 
\begin{equation}
    y_n\sim \mathcal T_{2\alpha}\Big(\mathbf x_n^\text T \boldsymbol\mu_\beta,  \frac{\gamma_*}{\alpha_*}  \big[\mathbf 1 + \mathbf x_\text{new}\boldsymbol\Lambda_*\beta\mathbf x_\text{new}^\text T\big]\Big)
\end{equation}
To come full-circle, notice that for $\boldsymbol\mu_0 = \mathbf 0$ and $\boldsymbol\Lambda_0 = \lambda^{-1}\mathbf I$ our estimate for $\boldsymbol\mu_*$ is ridge-optimal : 
\begin{equation}
    \boldsymbol\mu_* = \big(\mathbf X^\text T \mathbf X + \lambda\mathbf I\big)^{-1}\big(\mathbf X^\text T\mathbf y\big) = \hat{\boldsymbol\beta}_\text{RLS}
\end{equation}
Setting $\boldsymbol\mu_0 = \mathbf 0$ and $\boldsymbol\Lambda_0 = \lambda\mathbf I$ is a good place to start, but we'll be able to construct a far more interesting model by leveraging the fact that our data is temporal in nature. Bonds are constantly trading, and it is reasonable to assume that a recently estimated discount function is a reasonable prior for the current discount function. We can formalize this with the Kalman filter. 

\textbf{Sidebar :} For the reader unfamiliar with Bayesian learning, we can alternatively contextualize this process through the lens of convex optimization. For the conjugate Bayesian regression, the log-posterior is simply the sum of the log-likelihood (concave) and the log-prior (concave). The sum of concave functions is concave, and our parameters are defined over a convex set. Thus, conjugate posterior inference can be understood as unconstrained convex optimization.

\subsection{State-Space $\boldsymbol\beta$ with Kalman Filter} \label{kalman_filter}

We posit that over a set of states $\mathcal S$ we observe a stochastic parameter process for $\boldsymbol\beta$ : 
\begin{equation}
    \big\{\boldsymbol\beta_s\big\}_\mathcal S = \big\{\boldsymbol\beta_1, \boldsymbol\beta_2, \dots, \boldsymbol\beta_S\big\}
\end{equation}
In plain English, we're simply assuming that for an issuer, its discount function \emph{evolves over time} as real-world events impact investor expectations of the future, which are reflected in traded bond prices. We model this parameter process as a \textbf{Discrete Brownian Motion}, which is both a \emph{Markov process} (making it memoryless) and a \emph{martingale} (making it trendless). In general, this is one of the simplest stochastic processes that we are able to construct, though it is nonetheless still useful : 
\begin{align}
    \mathbb E\big[\boldsymbol\beta_{s}\mid \boldsymbol\beta_1,\boldsymbol\beta_2, \dots, \boldsymbol\beta_{s-1}\big] = \boldsymbol\beta_{s-1}
\end{align}
This implies that \emph{our best prediction for tomorrow's yield curve is today's yield curve with symmetric noise}. 

The natural statistical approach for modeling a discrete Brownian motion is the zero-drift \textbf{Kalman Filter}, which recursively models our parameter $\boldsymbol\beta$ with the following structure, similar models in \cite{dynamic_NS, dynamic_term_structure, jasic2018bayesian}. : 
\vspace{-0.3cm}
\begin{itemize}
    \item \textbf{Data Model} : $\mathbf p_s = \mathbf B_s \boldsymbol\beta_s + \boldsymbol\varepsilon_s$ (simply, the Vasicek model)
    \item \textbf{State Model} : $\boldsymbol\beta_s = \boldsymbol\beta_{s-1} + \boldsymbol\Delta_s$ with $\boldsymbol\Delta_s \sim \mathcal N\big(\boldsymbol\Delta_s \mid \mathbf 0, \boldsymbol\Sigma_s\big)$
\end{itemize}
\vspace{-0.3cm}
You'll notice that the State Model is an AR(1) model. 
Here, we introduce $\boldsymbol\Delta_s$ as the \emph{temporal change} of $\boldsymbol\beta$, with no drift and uncertainty $\boldsymbol\Sigma_s$. Since both $\boldsymbol\beta_{s-1}$ and $\boldsymbol\Delta_s$ follow Gaussian distributions, which is closed under addition, we obtain the following prior on $\boldsymbol\beta_s$ : 
\begin{equation}
    \mathbb P\big(\boldsymbol\beta_s\mid \sigma^2_s\big) = \mathcal N\big(\boldsymbol\beta_s \mid \boldsymbol\mu^*_{s-1},\sigma^2_s\boldsymbol\Lambda_{s-1}^* + \boldsymbol\Sigma_s\big)
\end{equation}
We can obtain a very convenient algebraic result by setting : 
\begin{align}
    \boldsymbol\Sigma_s = \delta_s^2 \cdot \boldsymbol\Lambda_{s-1}^*
\end{align}
The interpretation of this is that high-uncertainty parameters are more likely to evolve than low-certainty parameters, with $\delta_s^2$ acting as a \emph{variance amplifier} term, which is a hyperparameter specified by the user. Thus, our distribution now becomes : 
\begin{equation}
    \mathbb P\big(\boldsymbol\beta_s\mid \sigma^2_s\big) = \mathcal N\big(\boldsymbol\beta_s \mid \boldsymbol\mu^*_{s-1},\big[\sigma^2_s + \delta_s^2\big]\boldsymbol\Lambda_{s-1}^*\big)
\end{equation}
Since we treat $\sigma^2$ as an unknown in our Bayesian conjugate regression, we require one more step to make this all practical. We instead encode this uncertainty in the prior on $\sigma^2$ by constructing the following new random variable : 
\begin{equation}
    \tilde{\sigma}^2_s = \sigma_s^2 + \delta_s^2
\end{equation}
Thus we specify the full prior for our Bayesian model as : 
\begin{align}
    \mathbb P\big(\boldsymbol\beta_s, \tilde\sigma_s^2\big) &= \mathbb P\big(\boldsymbol\beta\mid \tilde\sigma^2\big) \cdot \mathbb P\big(\tilde\sigma^2\big)
    \\[5pt]
    \mathbb P\big(\boldsymbol\beta_s\mid \sigma^2_s\big) &= \mathcal N\big(\boldsymbol\beta_s \mid \boldsymbol\mu^*_{s-1},\tilde{\sigma}^2_s\cdot \boldsymbol\Lambda_{s-1}^*\big)
    \\[5pt]
    \mathbb P\big(\tilde\sigma^2_s\big) &= \mathcal G^{-1}\big(\tilde\sigma^2_s\mid\tilde\alpha, \tilde\gamma\big)
\end{align}
Where we have the following hyperparameters (see Appendix \ref{derivation_of_invgamma}) : 
\vspace{-0.3cm}
\begin{itemize}
    \item $\tilde \alpha_s  = \gamma_{s-1}^{*-1} \cdot (\alpha^*_{s-1} -1) \cdot (\alpha^*_{s-1} - 2) \cdot \big[(\gamma^*_{s-1} \cdot (\alpha^*_{s-1} -1)^{-1}) + \delta_s^2\big] + 2$
    \item $\tilde\gamma_s  = \big[\gamma^*_{s-1} \cdot (\alpha^*_{s-1} - 1)^{-1} + \delta^2_s\big]\cdot (\tilde\alpha_s - 1)^{-1}$
\end{itemize}
\vspace{-0.3cm}
And thus we have fully specified the model our state-space Bayesian approach for fitting the Vasicek exponential spline to corporate bonds in order to estimate a firm's default spread. 

\newpage
\section{Application}

As an illustrative example of our model's ability to anticipate a default, consider Diamond Offshore Drilling, a firm that filed for bankruptcy in 2021. Below we plot the 95\% confidence internal for this firm's 5-year model-estimated default spread (red), which we see steadily increases up until the bankruptcy even occurs. As a baseline to compare to, we also plot Amazon (green). 

\begin{figure}[h]
\includegraphics[scale=.7]{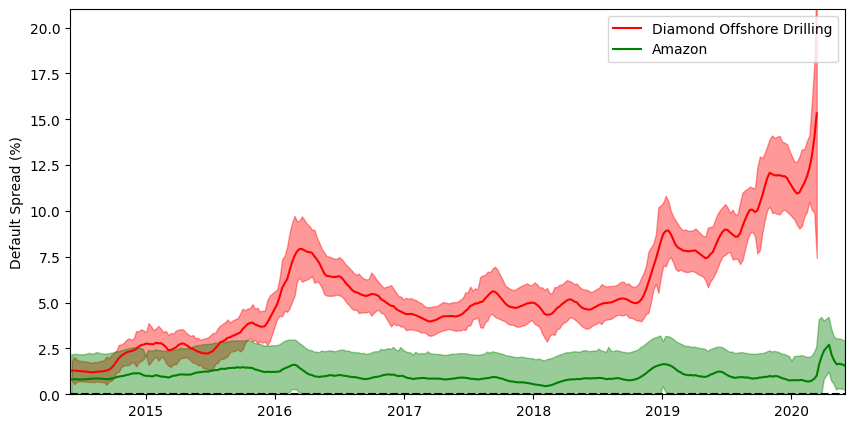}
\centering
\caption{5-Year Credit Spread for Diamond Offshore Drilling and Amazon}
\end{figure}

As a broader example of how our model is useful for portfolio construction, we compare the top and bottom quintiles of the 500 largest firms in the Russell 3000\footnote{Corporate bonds are most frequently issued by large cap firms, so we constrain our analysis to only this subset.} sorted by their model-estimated default spreads. 

\begin{figure}[h]
\includegraphics[scale=.53]{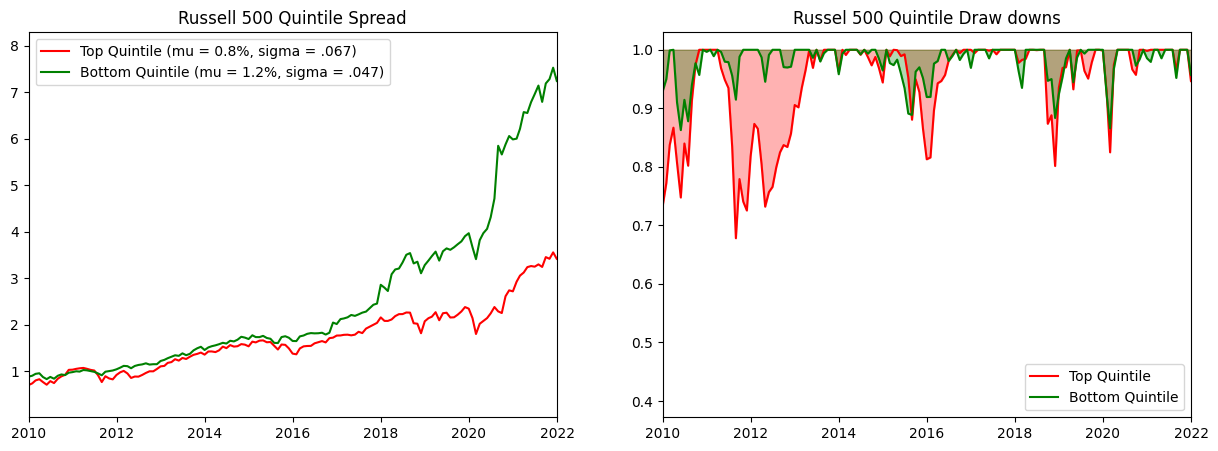}
\centering
\caption{Drawdowns and Quintile Spreads for R500}
\end{figure}
You'll notice that the portfolio of firms with highest default risk has lower performance and higher drawdowns, demonstrating that this signal effectively filters out many of the firms that have highly negative returns. Our sample size is small due to data limitations, but hopefully we can revisit this analysis in several years with more data to validate the robustness of this method. 

\newpage
\section{Conclusion}
In this paper demonstrate a novel way to estimate corporate yield curves that provides a \emph{high frequency} default risk signal. We accomplish this by adapting the Vasicek model to a Bayesian framework, refining a classic methodology in a way that is highly useful to investors. Our sample size is small, so this paper is more of a "proof of concept" rather than a highly reputable signal, but we hope with future iteration to improve upon it and expand the scope of its utility. 

\bibliographystyle{ieeetr}
\bibliography{main} 

\newpage
\appendix
\section{Constrained Regression with $\|\boldsymbol\beta\|_1 = 1$}\label{constrained_reg_proof}
Let $\mathbf X_k$ denote the $k$-th column of $\mathbf X \in \mathbb R^{N \times D}$. Construct : 
    \begin{align}
        \tilde{\mathbf y} &= \big[\mathbf y - \mathbf X_K\big] \in \mathbb R^N
        \\[5pt]
        \tilde{\mathbf X} &= \big[\mathbf X_1 - \mathbf X_K, \dots, \mathbf X_{K-1} - \mathbf X_K\big] \in \mathbb R^{N \times (K-1)}
    \end{align}
\textbf{(Lemma)} : The following two regressions have identical optimal $\boldsymbol\beta$ solutions : 
\vspace{-0.3cm}
\begin{itemize}
    \item \emph{Constrained Problem in $\mathbb R^K$} : $\mathbf y = \sum_{k=1}^K \mathbf X_k \beta_k + \boldsymbol\varepsilon$ subject to $1=\sum_{k=1}^K \beta_k$
    \item \emph{Unconstrained Problem in $\mathbb R^{K-1}$} : $\tilde{\mathbf y} = \sum_{k=1}^{K-1}\tilde{\mathbf X}_k\beta_k + \boldsymbol\varepsilon$
\end{itemize}
\vspace{-0.3cm}
\textbf{(Proof)} We'll start off with the first regression, and will derive from it the second : 
\begin{align}
    \mathbf y&= \sum_{k=1}^{K-1} \mathbf X_k\beta_k + \mathbf X_K\beta_K+ \boldsymbol\varepsilon
    \\
    \mathbf y&= \sum_{k=1}^{K-1}\mathbf X_k \beta_k + \mathbf X_K\bigg(1 - \sum_{i=1}^{K-1}\beta_k\bigg)+ \boldsymbol\varepsilon
    \\
    \mathbf y&= \sum_{k=1}^{K-1}\big(\mathbf X_k - \mathbf X_K\big)\beta_k  + \mathbf X_K + \boldsymbol\varepsilon
    \\
    \mathbf y - \mathbf X_K &=\sum_{k=1}^{K-1}\big(\mathbf X_k - \mathbf X_K\big)\beta_k    + \boldsymbol\varepsilon
    \\
    \tilde{\mathbf y} &= \sum_{k=1}^{K-1}\tilde{\mathbf X}_k\beta_k + \boldsymbol\varepsilon 
\end{align}

\vspace{-1cm}
\section{Parameters for Shifted Inverse Gamma Distribution}\label{derivation_of_invgamma}

Consider $\sigma^2 \sim \mathcal G^{-1}(\alpha,\gamma)$ with $\mathbb E\big[\sigma^2 \big] = \frac{\gamma}{\alpha-1}$ for $\alpha > 1$ and $\mathbb V\big[\sigma^2 \big] = \frac{\gamma^2}{(\alpha-1)^2(\alpha-2)} = \frac{\mathbb E[\sigma^2 ]^2}{(\alpha-2)}$ for $\alpha > 2 $

Define new variable $\tilde \sigma^2  \sim \mathcal G^{-1}(\tilde\alpha, \tilde\gamma)$ such that $\mathbb V\big[\tilde \sigma^2 \big] = \mathbb V\big[\sigma^2 \big]$ and $\mathbb E\big[\tilde \sigma^2 \big] = \mathbb E\big[\sigma^2 \big] + \delta^2$.
Define : 
\vspace{-0.3cm}
\begin{itemize}
    \item $\tilde E = \mathbb E\big[\sigma^2\big] + \delta^2 = \gamma \cdot (\alpha - 1)^{-1} + \delta^2$
    \item $\tilde V = \mathbb V\big[\sigma^2\big] + \epsilon = \tilde E^2 \cdot (\alpha - 2) + \epsilon$
\end{itemize}
\vspace{-0.3cm}
We solve for $\tilde\alpha$ as : 
\begin{align}
    \mathbb V\big[\tilde\sigma^2\big] &= \tilde V \\
    \tilde E^2 \cdot (\tilde\alpha - 2)^{-1} &= \tilde V \\
    \tilde \alpha &= \tilde E^2 \cdot \tilde V^{-1} + 2
\end{align}
We then solve for $\tilde\gamma$ as : 
\begin{align}
    \mathbb E\big[\tilde\sigma^2\big] &= \tilde E \\
    \tilde\gamma \cdot (\tilde\alpha -1 )^{-1} &= \tilde E \\
    \tilde\gamma &= \tilde E \cdot (\tilde\alpha - 1)
\end{align}

\section{Model Stability Notes}
\subsection{Variance Amplifier Hyper parameters}
As mentioned in Section \ref{kalman_filter} we add a \emph{variance amplifier} hyper parameter $\delta^2 > 0$ to the prior on the coefficients that describe the yield curve. This is to combat the issue known as vanishing variance. This is problematic for model stability, because as variance tends toward zero the new calculated state will always be the same as the previous state.

This phenomenon tends to plague Markov models that model the new states variance as a function of the previous states parameters. However in this model we also model the \emph{variance of the variance} as a function of the previous states parameters (see equations \ref{eq_47} and \ref{eq_48} for how $\alpha$ and $\gamma$ relate to the previous state)

This again is problematic as if VOV tends to zero then our previously defined hyper paramter $\delta^2$ will ensure that our variance grows monotonically and infinitely. This will lead to overinflated and practically unusable confidence intervals as well as unstable coefficients. To address this we add a second hyper paramer $\epsilon > 0$ to the new states estimate of VOV. See Section \ref{derivation_of_invgamma} for algebraic details.

\subsection{Rank Deficient Matrices}
The user may encounter \emph{rank deficient} matrices when using equation \ref{eq_45} to compute the covariance structure of the coefficients. In particular this issue will almost certaintly arise after the first iteration where the prior $\mathbf \Lambda_0$ is no longer uninformative and is instead the previous states posterior. This is due to a well known result in linear algebra: \emph{The Rank Product Theorem}
\begin{align}
    \text{Rank}(AB) \leq \text{min}(\text{Rank}(A), \text{Rank}(B))
\end{align}
As many firms issue fewer bonds than $K$, from the theorem above one can verify that:
\begin{align}
    \text{Rank}({\mathbf{X^TX}}) < K
\end{align} 

Thus $\mathbf{X^TX}$ is \emph{rank deficient} and hence non-invertible. In the initial iteration equation \ref{eq_45} is solved by setting  $\mathbf{\Lambda_*} = \lambda \mathbf I$ (the ridge optimal solution). While this solution will work for all states of the model we can still impose a prior on the covariance structure between coefficients by setting $\mathbf{\Lambda_*} = \mathbf{\Lambda_0} + \lambda \mathbf I$ for further iterations. This can be viewed as weakening the prior on the covariance matrix and is similar in spirit to our variance multiplier hyperparameters.

\end{document}